\documentclass[twocolumn,secnumarabic,amssymb, nobibnotes, aps, showpacs, superscriptaddress]{revtex4}

\usepackage{graphicx}
\usepackage{dcolumn}
\usepackage{bm}
\usepackage{color} 
\begin{document}


\title{Coexistence of electron and hole transport in graphene}

\author{S.~Wiedmann}
\affiliation{Radboud University Nijmegen, Institute for Molecules and Materials and High Field Magnet Laboratory,
Toernooiveld 7, 6525 ED Nijmegen, The Netherlands}

\author{H.~J.~van Elferen}
\affiliation{Radboud University Nijmegen, Institute for Molecules and Materials and High Field Magnet Laboratory,
Toernooiveld 7, 6525 ED Nijmegen, The Netherlands} 

\author{E.~V.~Kurganova}
\affiliation{Radboud University Nijmegen, Institute for Molecules and Materials and High Field Magnet Laboratory,
Toernooiveld 7, 6525 ED Nijmegen, The Netherlands}

\author{M.~I.~Katsnelson}
\affiliation{Radboud University Nijmegen, Institute for Molecules and Materials, Heyendaalseweg 135, 6525 AJ Nijmegen,
The Netherlands}

\author{A.~J~.M.~Giesbers}
\affiliation{Max-Planck-Institut f\"ur Festk\"orperforschung, Heisenbergstra\ss e 1, 70569 Stuttgart, Germany}

\author{A.~Veligura}
\affiliation{Physics of Nanodevices, Zernike Institute for Advanced Materials, 
University of Groningen, Nijenborgh 4, 9747 AG Groningen, The Netherlands}

\author{B.~J.~van Wees}
\affiliation{Physics of Nanodevices, Zernike Institute for Advanced Materials, 
University of Groningen, Nijenborgh 4, 9747 AG Groningen, The Netherlands}

\author{R.~V.~Gorbachev}
\affiliation{Department of Physics, University of Manchester, M13 9PL Manchester, United Kingdom}

\author{K.~S.~Novoselov}
\affiliation{Department of Physics, University of Manchester, M13 9PL Manchester, United Kingdom}

\author{J.~C.~Maan}
\affiliation{Radboud University Nijmegen, Institute for Molecules and Materials and High Field Magnet Laboratory,
Toernooiveld 7, 6525 ED Nijmegen, The Netherlands}

\author{U.~Zeitler}
\affiliation{Radboud University Nijmegen, Institute for Molecules and Materials and High Field Magnet Laboratory,
Toernooiveld 7, 6525 ED Nijmegen, The Netherlands}

\date{\today}

\begin{abstract}
When sweeping the carrier concentration in monolayer graphene through 
the charge neutrality point, the experimentally measured Hall resistivity 
shows a smooth zero crossing. Using a two-component model of coexisting 
electrons and holes around the charge neutrality point, we unambiguously 
show that both types of carriers are simultaneously present. For high 
magnetic fields up to 30~T the electron and hole concentrations at the 
charge neutrality point increase with the degeneracy of the zero-energy 
Landau level, which implies a quantum Hall metal state at $\nu$=0 made 
up by both electrons and holes. 
\end{abstract}


\maketitle

\section{Introduction}

The carrier concentration in semiconductors is commonly measured using 
the Hall effect based on the Lorentz force exerted on moving charged 
particles in a perpendicular magnetic field \cite{Hall}. In conventional 
finite-gap semiconductors, the low-temperature Hall resistivity $\rho_{xy}$ 
directly measures either the electron or the hole density. However, in 
compensated semiconductors, where electrons and holes coexist, the Hall 
resistivity is determined by {\sl both} types of carriers and, 
in particular, becomes zero in a fully compensated material. 

Graphene is an ideal two-dimensional zero-gap semiconductor with a linear 
dispersion \cite{Rise of graphene} where the electron and hole concentration 
at $T$=0 go to zero when sweeping the carrier density through the charge 
neutrality point (CNP). However, non-perfect samples with random potential 
fluctuations will break up into spatially inhomogeneous conducting electron-hole 
puddles \cite{Yacobi-puddles} leaving a finite number of electrons and holes 
directly at the CNP. 

In this article, we present experimental results on the Hall resistivity 
$\rho_{xy}$ in graphene around the CNP in magnetic fields up to 30~T 
and for temperatures down to 0.5~K. We demonstrate that the smooth 
zero-crossing of $\rho_{xy}$ at the CNP for all magnetic fields is caused 
by a finite concentration of both electrons and holes below and above the 
CNP with an equal number of electron and hole states occupied at the CNP. 
We show that the measured carrier concentration increases linearly with 
the magnetic field which is related to the degeneracy of the zero-energy 
LL shared equally between electrons and holes.

We have investigated three different graphene devices made from Kish 
graphite (sample A) and natural graphite (samples B and C) with 
mobilities between $\mu=0.8$~m$^2$V$^{-1}$s$^{-1}$ for sample A and 
$\mu=1$~m$^2$V$^{-1}$s$^{-1}$ for samples B and C. Single-layer graphene 
flakes were deposited on a Si/SiO$_{2}$ substrate, identified optically~\cite{Blake} 
and patterned using standard techniques \cite{ScienceKostya-Device, Rise of graphene}. 
The total charge-carrier concentration $q$ in the graphene films, defined as 
$q \equiv n-p\simeq \alpha V_{g}$, can be adjusted from hole-doped ($q<0$) 
to electron doped ($q>0$) by means of a back-gate voltage $V_{g}$. Here $n$ and 
$p$ are the carrier concentrations for electrons and holes, respectively, and 
$\alpha=7.2\times 10^{14}$~m$^{-2~}$V$^{-1}$ for a 300-nm thick SiO$_2$ gate 
insulator. In order to remove surface impurities, all devices were annealed 
at 440~K prior to the low-temperature measurements. Admixtures of $\rho_{xx}$ 
to $\rho_{xy}$ due to contact misalignment and inhomogeneities we removed by 
symmetrization of all traces measured in positive and negative magnetic 
fields.

The paper is organized as follows: Section II presents our experimental results.
The first part of Section II shows transport measurements at low magnetic fields 
where the Hall resistance is not yet quantized and charge carriers can be
considered as ``free" (mobile). The second part presents data up to 30~T in  
the quantum Hall (QH) regime. Section III develops a model for the density
of states in graphene, first applied to our samples and then we discuss
different splitting scenarios of the lowest Landau level. Concluding remarks are 
given in the last section.

\section{Experimental results}

\begin{figure}[t]
\includegraphics[width=9cm]{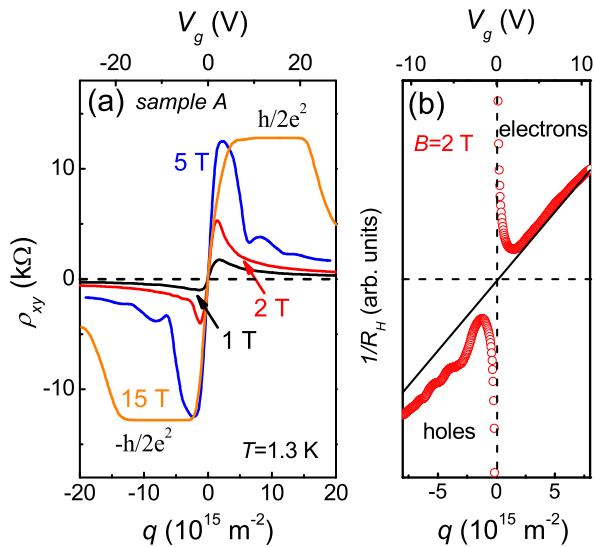}
\caption{\label{fig1} (Color online) (a) Dependence of $\rho_{xy}$ in sample A 
         on the carrier concentration (bottom axis) or on the back-gate voltage 
         $V_{g}$ (top axis) for several magnetic fields at $T$=1.3~K. (b) Inverse 
         Hall coefficient $1/R_{H}$ as a function of $q$ for $B$=2~T. The solid 
         line shows the expected behavior of a conventional zero-gap semiconductor 
         where electrons and holes get fully depleted at the CNP.}
\end{figure}

We first present measurements of Hall resistivity $\rho_{xy}$ with increasing 
magnetic field in Fig.~\ref{fig1}(a) for sample A as a function of total carrier 
concentration $q$ for several magnetic fields at $T$=1.3~K. The corresponding 
back-gate voltage $V_{g}$ is displayed on the top-axis. For $B$=15~T, $\rho_{xy}$ 
exhibits Hall plateaus quantized to $\rho_{xy}= \pm h/2e^{2}$ at filling factors 
$\nu=\pm 2$. For all magnetic fields, the Hall resistance is not diverging at the 
CNP when either electron or hole states are depleted. $\rho_{xy}$ rather moves 
smoothly through zero from the $\nu=-2$ plateau to the $\nu$=+2 plateau. 

In order to accommodate for this simple experimental observation, we 
describe the inverse Hall coefficient $1/R_H = B/\rho_{xy}$ with a two-carrier model for electrons and holes as known for compensated 
semiconductors \cite{two-carrier}
\begin{equation}
    \frac{1}{R_{H}} = \frac{e(n\mu_{n}+p\mu_{p})^{2}}{ n\mu_{n}^{2}-p\mu_{p}^{2}}. 
    \label{eq1}
\end{equation}
$n$ and $p$ are the electron and hole concentrations and $\mu_e$ and 
$\mu_h$ are the electron and hole mobilities, respectively. In our 
graphene samples the measured conductivity as a function of carrier 
concentration is symmetric around the CNP and we can therefore assume 
the same mobility for both electrons and holes, $\mu_{n}=\mu_{p}$, 
and Eq.~(\ref{eq1}) simplifies to
\begin{equation}
    \frac{1}{R_{H}} = \frac{e(n+p)^{2}}{(n-p)}. 
    \label{eq2}
\end{equation}
It is worth emphasizing that we can apply the two-carrier model despite 
the presence of electron-hole puddles, which would result, for conventional 
nonrelativistic charge carriers, in spatial separation and related percolation 
phenomena in electron and hole regions. In the two-dimensional case, 
the percolation over electron puddles blocks unavoidably the transport for 
holes, and vice versa. The case of graphene is dramatically different.
The crucial point is that for graphene the borders between $p$ and $n$ 
regions are actually transparent, and electrons and holes transfer smoothly 
into each other, which is referred to as Klein tunneling \cite{Katsnelson-Klein}. 
At specific magic angles of incidence (including normal incidence) the 
transmission probability is 100\%\ . The presence of a magnetic field does not 
destroy the Klein tunneling but just shifts the magic angles \cite{Young-Klein}.
It can be assumed that tunneling from one electron puddle to the other electron 
puddle always remains possible even for carriers incident to an oblique angle
(the same holds for hole transport). Thus, even under a nonuniform distribution
of electron-hole puddles, we can apply our two-carrier model to graphene.

Fig. \ref{fig1}(b) shows the inverse Hall coefficient $1/R_{H}$ as a 
function of $q$ for $B$=2~T extracted from our measurements. For high 
$q$, $1/R_{H}$ exhibits a linear increase due to the presence of 
either electrons ($q>0$) or holes ($q<0$). For $q \rightarrow 0$, 
however, the simultaneous presence of two distinct types of charge carriers around 
the CNP immediately becomes visible as a divergence of $1/R_{H}$ 
around the CNP, which in turn implies that $n+p$ must remain finite. 

\subsection{Low magnetic fields}

\begin{figure}[t]
\includegraphics[width=9cm]{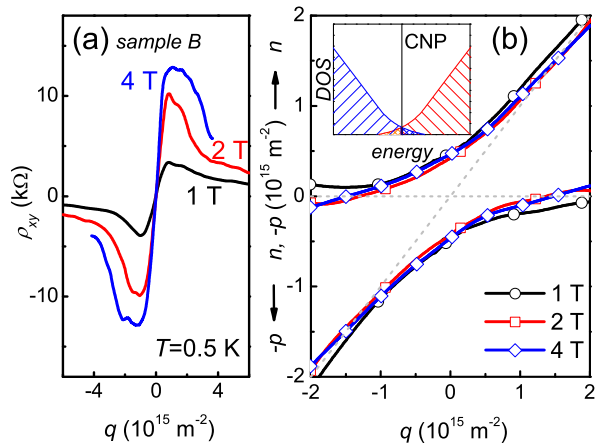}
\caption{\label{fig2} (Color online) (a) Low-field Hall resistivity $\rho_{xy}$ and (b) extracted 
         carrier concentration for electrons $n$ and holes $p$ as a function of total charge $q$ 
         according to Eq.~(\ref{eq2}) for sample B. Both types of charge carriers are present for 
         $|q|<2\cdot10^{15}$~m$^{-2}$. Inset: Sketch of the DOS for $B$=0 at the CNP.}
\end{figure}

We now present low-field data in Fig.~\ref{fig2} for sample B measured at 0.5~K 
and in magnetic fields where the quantum Hall effect (QHE) is not yet developed. Using Eq.~(\ref{eq2}) we 
extract the individual charge-carrier concentrations $n$ and $p$ as a function 
of the total charge density $q$ [see Fig.~\ref{fig2}(b)].
Both charge carriers are present above and below the CNP and the electron (hole) 
concentration already starts to increase as the hole (electron) concentration is still 
decreasing. Precisely at the CNP, we extract a charge-carrier concentration 
$n(q=0)=p(q=0) = 4.2\times 10^{14}$~m$^{-2}$ only weakly dependent on $B$ for 
$0 < B < 4$~T. Away from the CNP, the system remains two-component and the minority 
charge carriers only disappear for $|q|>2 \times 10^{15}$~m$^{-2}$. The same analysis 
for the other two samples qualitatively yields similar results with 
$n(q=0)=p(q=0) = 7.4 \times 10^{14}$~m$^{-2}$ for sample A and 
$n(q=0)=p(q=0) \simeq 5 \cdot 10^{14}$~m$^{-2}$ for sample C. 
The fact that the sample with the lowest mobility (sample A) reveals the 
highest $n(q=0)$ qualitatively confirms a scenario of coexisting electron-hole 
puddles, where lower mobilities are generally associated with larger potential 
fluctuations. 

\subsection{Quantum Hall regime}

\begin{figure}[t]
\includegraphics[width=9cm]{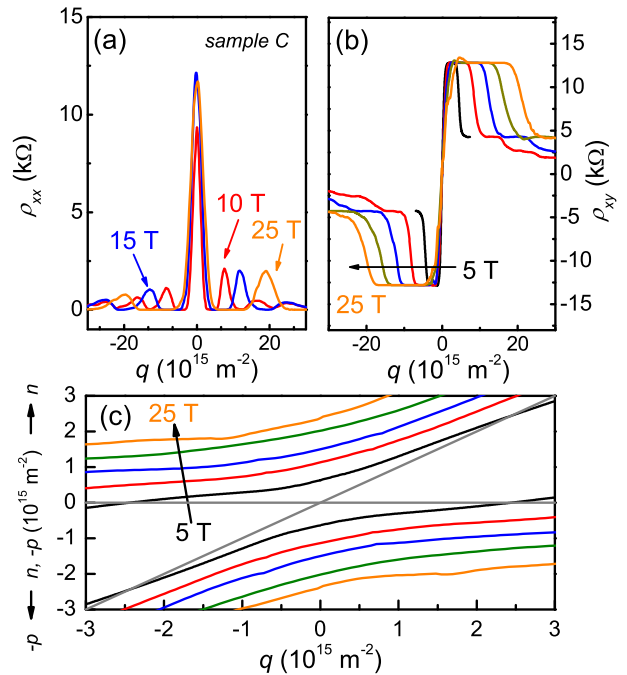}
\caption{\label{fig3} (Color online) (a) High-field Hall resistivity $\rho_{xy}$
         and longitudinal resistivity $\rho_{xx}$ of sample C at 4~K for different
         magnetic fields. (c) Corresponding charge-carrier concentrations 
         $n$ and $p$ extracted according to Eq.~(\ref{eq2}).}
\end{figure}

We now turn our attention to measurements in high magnetic fields. We present
experimental data from 5 to 25~T both for longitudinal resistivity $\rho_{xx}$ and 
Hall resistivity $\rho_{xy}$ [see Fig.~\ref{fig3}(a) and Fig.~\ref{fig3}(b)] measured in sample C at $T$=4~K. 
$\rho_{xy}$ is measured from $B$=5~T up to 25~T in steps of 5~T. $\rho_{xy}$~is 
now quantized at $\nu=\pm$2 but still shows a smooth zero-crossing from 
$\rho_{xy} = -h/2e^2$ to $\rho_{xy}=+h/2e^2$ without any sign of divergence 
at the CNP. Consequently, we still find a finite charge carrier concentration 
for electrons and holes around the CNP as depicted in Fig.~\ref{fig3}(c). 
Therefore, we can conclude that electrons still contribute to conduction 
below $E$=0 and holes do so above $E$=0.
It should be noticed that in the range of magnetic fields used for the extraction
of the carrier densities, $\rho_{xx}$ does not affect $\rho_{xy}$ even if we take
into account a small amount of mixing between both signals but becomes relevant
if $\rho_{xx}$ starts diverging at the CNP. 

\begin{figure}[t]
\includegraphics[width=9cm]{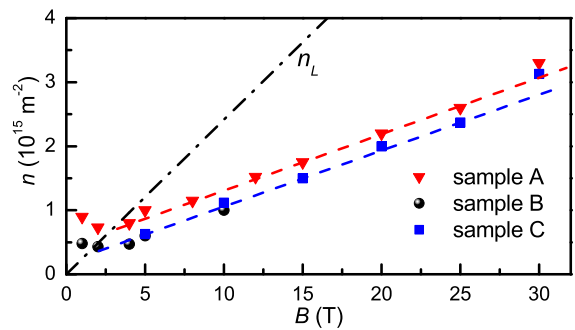}
\caption{\label{fig4} (Color online) Charge-carrier concentration at the 
         CNP as a function of magnetic field for all samples (the dashed-dotted 
         line depicts degeneracy $n_{L}$).}
\end{figure}

In addition, we now observe an increase of $n$ and $p$ with increasing 
magnetic field. This field-dependent carrier concentration around the CNP 
is elucidated further in Fig.~\ref{fig4} where we plot the electron 
concentration at the CNP as a function of magnetic field for all investigated 
samples. For low magnetic fields ($B<$5~T), $n(q=0)$ remains constant 
and can be explained by the presence of electron-hole puddles. 

For higher magnetic fields, $n(q=0)$ starts to increase linearly 
with $B$ reflecting the $B$-proportional degeneracy $n_{L}$ for 
each of the four sub-levels in the zero-energy Landau level (LL) 
\cite{Kostya-QHE, Kim-QHE}. Since, at the CNP, half of the possible 
electrons states and half of the hole states are filled, respectively, we 
expect $n_{L}$ electron states and $n_{L}$ hole states occupied per 
unit area. Therefore, for comparison, we have also plotted $n_{L}$ in 
Fig.~\ref{fig4}. Interestingly, we only observe about 30~$\%$ 
of the expected electron and hole concentration $n_{L}$ in our data 
extracted from the Hall experiments.

In contrast to low magnetic fields, where the QHE is not yet developped
and all charge carriers can be considered as mobile, we now have to
take into account localized charge carriers in the tails of the LLs in the 
quantized regime. This fact is essentially reflected in Fig.~\ref{fig4}, 
where we extracted the density for electrons and holes using Eq.~(\ref{eq2}) 
from the Hall resistivity, which only takes free charge carriers into 
account. To further support our assumption that about 30~$\%$ of the charge carriers
are indeed free, we take a look at the broadening of LLs and the ratio between extended 
and localized states depending on the strength of the magnetic field which
has been extracted from temperature-dependent measurements \cite{Jos-temperature} 
and QH plateaus at high $B$. Indeed, we observe a good agreement with our findings 
that only 30~$\%$ of the total carrier concentration is measured as free 
charge carriers. 

\section{Density of states model}

\subsection{Investigated samples}

\begin{figure}[t]
\includegraphics[width=9cm]{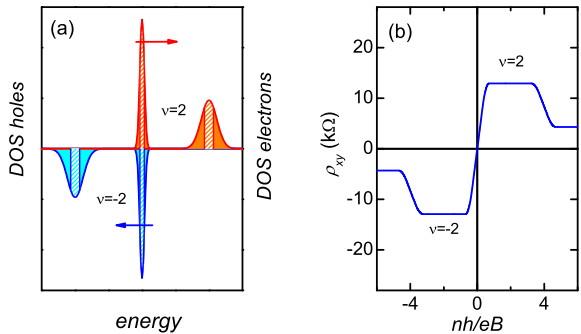}
\caption{\label{fig5} (Color online) (a) Four-fold degenerate zeroth LL 
         with coexisting electrons and holes below and above the CNP 
         (the electron levels (red) are sketched upward and the hole levels
         (blue) are sketched downward). (b) Smooth zero-crossing of the 
         Hall resistance from $\nu$=-2 to $\nu$=2.}
\end{figure}

The above measurements allow us to sketch the density of states (DOS) for 
electrons and holes. For $B$=0 (see inset to Fig.~\ref{fig2}) the DOS 
in graphene $D(E)=2|E|/\pi(\hbar v)^{2}$ ($v$ is the Fermi velocity) is 
smeared out around the CNP due to the presence of electron-hole puddles. 
Applying a magnetic field leads to a quantization of the DOS, shown in 
Fig. \ref{fig5}(a). Electrons and holes in the center of the LLs are 
extended (shaded areas) whereas they are localized in the Landau level 
tails (filled areas). In that picture the LLs $N$=0 and $N$=1 are well 
separated, yielding quantized plateaus in $\rho_{xy}$ at $\nu=\pm$2 
[Fig.~\ref{fig3}(a)] when the Fermi energy is situated in the localized 
tails of the LLs. 

Within this DOS model (see also Ref. \cite{Jos-gap}) we can now calculate 
the longitudinal conductivity $\sigma_{xx}$ by means 
of the Kubo-Greenwood formalism \cite{Kubo, Greenwood} 
and the Hall conductivity $\sigma_{xy}$ summing up all states below 
the Fermi energy \cite{Streda}. Including the presence of electrons and 
holes above and below the CNP indeed yields a smooth zero crossing of 
$\rho_{xy}$ as measured in Fig. \ref{fig3}(a) and modeled in Fig. \ref{fig5}(b).

Our experimental observation of coexisting electrons and holes 
around the CNP also has a direct implication on the nature of 
the $\nu$=0 QHE in graphene~\cite{DasSarma}. Neither a gap opening at 
the CNP \cite{Jos-gap}, nor a complete lifting of spin and valley 
degeneracy, if we assume the spin first scenario of the zeroth LL \cite{Zhang}, 
fundamentally change the zero-crossing of the Hall resistance. Our experimental 
results up to a magnetic field of 30~T do not exhibit interaction-driven 
QHE \cite{Nomura} due to a larger disorder confirmed by lower 
mobility in our samples compared to Ref. \cite{Zhang}. If we calculate 
$\sigma_{xy}$ and $\sigma_{xx}$ from measured $\rho_{xy}$ and 
$\rho_{xx}$ using standard matrix inversion our samples show the gap 
opening in $\sigma_{xx}$ at $B$=30~T due to increasing $\rho_{xx}$ 
at the CNP. Consequently, a small plateau in $\sigma_{xy}$ at the CNP 
appears whereas the Hall resistance smoothly crosses through zero 
\cite{Jos-gap}. 

\subsection{Splitting scenarios of the lowest Landau level}

\begin{figure}[t]
\includegraphics[width=9cm]{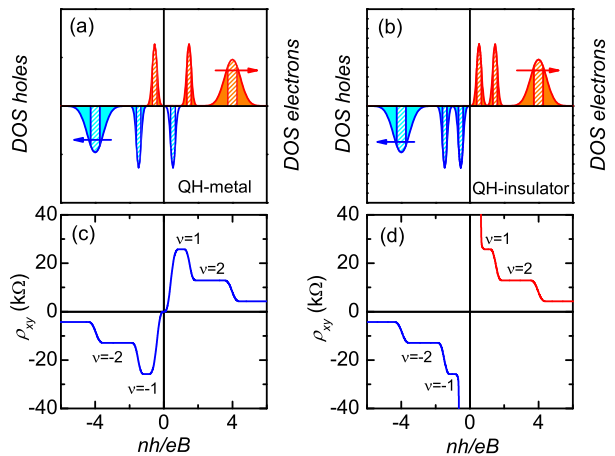}
\caption{\label{fig6} (Color online) 
Sketched DOS for (a) a QH-metal and (b) a QH-insulator if both spin and
valley splitting is resolved. 
(c) Smooth zero-crossing of the Hall resistivity if both charge carriers are
present above and below the CNP and 
(d) diverging $\rho_{xy}$ due to electrons and holes being separated and residing on
different sides of the CNP.}
\end{figure}

In samples with lower disorder, spin and valley degeneracies are lifted. 
As demonstrated in Ref. \cite{Zhang}, the Hall resistivity exhibits a smooth 
zero-crossing (with fluctuations) from the $\nu$=-1 plateau to the $\nu$=1 
plateau with increasing $V_{g}$. We have calculated the DOS assuming that both
electrons and holes exist above and below the CNP [Fig. \ref{fig6}(c)] and
find the smooth zero-crossing of the Hall resistivity [see Fig. \ref{fig6}(c)].

Furthermore, we use our DOS model to directly address the question weather
$\nu$=0 is a QH-metal or a QH-insulator. Measurements of longitudinal
resistance have shown either finite $\rho_{xx}$, even subjected to high 
magnetic fields (QH-metal) \cite{Kostya-QHE, Kim-QHE, Zhang, Abanin}
or a steeply increase in $\rho_{xx}$, attributed to an insulating
ground state \cite{Ong}. The first observation is generally explained by an
insulating bulk and conducting channals at the sample edges \cite{Abanin}.  
Both scenarios are directly related to the lifting of degeneracy of the
zeroth LL. Whereas in a QH-metal spin splitting is larger than valley splitting, 
in a QH-insulator the contrary is the case. If spin and valley degeneracy is lifted, 
a zero-crossing of $\rho_{xy}$ is observed if we include the presence of
electrons and holes above and below the CNP, see Fig. \ref{fig6}(a) and (c). 
However, if we separate electrons and holes at the CNP, see Fig. \ref{fig6}(b)
(valley first scenario), $\rho_{xy}$ diverges (see Fig. \ref{fig6}(d)). 
The divergence of $\rho_{xy}$ in the valley first scenario
beyond filling factor $\nu$=1 when approaching the CNP has 
indeed been recently found in high mobility graphene devices, 
fabricated on a single-crystal boron nitride substrate \cite{Dean} and thus 
confirm our DOS model. However, beyond fractional filling factor $\nu$=1/3, $\rho_{xy}$ 
starts to decrease strongly and might pass through zero. This behavior would
imply positive and negative charged composite fermions around the
CNP.

\section{Conclusion}

In conclusion, we have performed measurements of the Hall resistivity in 
graphene in a magnetic field up to 30~T. $\rho_{xy}$ does not diverge 
at the CNP but shows a smooth transition from electrons to holes. Our 
analysis based on mixed conduction at the CNP implies that both electrons 
and holes exist both below and above the CNP with as many hole states 
as electron states occupied at the CNP. Charge-carrier concentration 
as a function of magnetic field is explained as a transition from 
transport dominated by electron-hole puddles to a quantized DOS 
with increasing $B$. Taking into account the presence of both charge 
carriers above and below the CNP contributes to a better understanding 
of the unique nature of electronic states at the lowest LL in graphene.
\\Finally, we have to point out that physics around the CNP, such as the
behavior of $\rho_{xy}$ from hole-dominated to electron-dominated transport
becomes easier to access with high-mobility samples even though diverging
$\rho_{xx}$ directly affects the extraction of Hall resistivity under realistic
experimental conditions.

Part of this work has been supported by EuroMagNET II under the EU 
contract number 228043 and by the Stichting Fundamenteel Onderzoek 
der Materie (FOM) with financial support from the Nederlandse
Organisatie voor Wetenschappelijk Onderzoek (NWO).


\begin{thebibliography}{22}

\bibitem{Hall}
E. H. Hall, American Journal of Mathematics vol \textbf{2}, 287 (1879).

\bibitem{Rise of graphene}
A. K. Geim and K. S. Novoselov, Nat. Mater. \textbf{6}, 183 (2007).

\bibitem{Yacobi-puddles}
J. Martin, N. Akerman, G. Ulbricht, T. Lohmann, J. H. Smet, K. von Klitzing, A. Yacobi, 
Nat. Phys. \textbf{4}, 144 (2008).

\bibitem{Blake}
P. Blake, E. W. Hill, A. H. Castro Neto, K. S. Novoselov, D. Jiang, R. Yang, T. J. Booth, and A. K. Geim, Appl. Phys. Lett. \textbf{91}, 063124 (2007).

\bibitem{ScienceKostya-Device}
K. S. Novoselov, A. K. Geim, S. V. Morozov, D. Jiang, Y. Zhang, S. V. Dubonos, I. V. Grigorieva, and A. A. Firsov, 
Science \textbf{306}, 666 (2004).

\bibitem{two-carrier}
see, e.g., K. Seeger, \textit{Semiconductor physics, an introduction}, 5th ed., p. 61, Springer, Berlin (1997).

\bibitem{Katsnelson-Klein}
M. I. Katsnelson, K. S. Novoselov, and A. K. Geim, Nature Phys. \textbf{2}, 620 (2006).

\bibitem{Young-Klein}
A. F. Young and P. Kim, Nature Phys. \textbf{5}, 222 (2009).

\bibitem{Kostya-QHE}
K. S. Novoselov, A. K. Geim, S. V. Morozov, D. Jiang, M. I. Katsnelson, I. V. Grigorieva, S. V. Dubonos, and A. A. Firsov, 
Nature (London) \textbf{438}, 197 (2005).

\bibitem{Kim-QHE}
Y. Zhang, Y. Tan, H. L. Stormer, and P. Kim, Nature (London) \textbf{438}, 201 (2005).

\bibitem{Jos-temperature}
A. J. M. Giesbers, U. Zeitler, M. I. Katsnelson, L. A. Ponomarenko, T. M. Mohiuddin, and J. C. Maan, 
Phys. Rev. Lett. \textbf{99}, 206803 (2007). 

\bibitem{Jos-gap}
A. J. M. Giesbers, L. A. Ponomarenko, K. S. Novoselov, A. K. Geim, M. I. Katsnelson, J. C. Maan, and U. Zeitler, 
Phys. Rev. B \textbf{80}, 201403(R) (2009).

\bibitem{Kubo}
R. Kubo, Canad. J. Phys. \textbf{34}, 1274-1277 (1956).

\bibitem{Greenwood}
D. A. Greenwood, Proc. Phys. Soc. London \textbf{71}, 585-596 (1958).

\bibitem{Streda}
P. Streda, J. Phys. C: Solid. State Phys. \textbf{15}, 717-721 (1982).

\bibitem{DasSarma}
S. Das Sarma and K. Yang, Solid State Commun. \textbf{149}, 1502 (2009).

\bibitem{Zhang}
Y. Zhang, Z. Jiang, J. P. Small, M. S. Purewal, Y. W. Tan, M. Fazlollahi, J. D. Chudow, J. A. Jaszczak, H. L. Stormer, and P. Kim, 
Phys. Rev. Lett. \textbf{96}, 136806 (2006).

\bibitem{Nomura}
K. Nomura and A. H. MacDonald, Phys. Rev. Lett. \textbf{96}, 256602 (2006).

\bibitem{Abanin}
D. A. Abanin, K. S. Novoselov, U. Zeitler, P. A. Lee, A. K. Geim, and L. S. Levitov, Phys. Rev. Lett. \textbf{98}, 196806 (2007).

\bibitem{Ong}
J. G. Checkelsky, L. Li, and N. P. Ong, Phys. Rev. Lett. \textbf{100}, 206801 (2008); Phys. Rev. B 
\textbf{79}, 115434 (2009).
        
\bibitem{Dean}
C. R. Dean, A. F. Young, P. Cadden-Zimansky, L. Wang, H. Ren, K. Watanabe, T. Taniguchi, P. Kim, J. Hone, K. L. Shepard,
Nature Phys., \textbf{5}, 693 (2011).
              
\end{thebibliography}
\end{document}